\documentclass[aps,prl,twocolumn,epsfig,preprintnumbers,superscriptaddress,10pt]{revtex4}

\usepackage[dvips]{color}
\usepackage[normalem]{ulem}

 \usepackage{epsfig}

\begin{document}

\title{Calculating the Jet Quenching Parameter from AdS/CFT}

\author{Hong Liu}
\affiliation{Center for Theoretical Physics, Massachusetts Institute
of Technology, Cambridge, MA 02139, USA}
\author{ Krishna Rajagopal} 
\affiliation{Center for Theoretical Physics, Massachusetts Institute
of Technology, Cambridge, MA 02139, USA}
\affiliation{Nuclear Science Division, MS 70R319,
Lawrence Berkeley National Laboratory, Berkeley, CA 94720, USA}
\author{Urs Achim Wiedemann}
\affiliation{Department of Physics and Astronomy, University of Stony Brook, NY 11794, USA}
\affiliation{RIKEN-BNL Research Center, Brookhaven National Laboratory, Upton, NY 11973-5000, USA}

\preprint{MIT-CTP-3739, RBRC-601}


\begin{abstract}
Models of medium-induced radiative parton energy loss account for the strong suppression of 
high-$p_T$ hadron spectra in $\sqrt{s_{NN}}=200$ GeV Au-Au collisions at RHIC
in terms of  a single ``jet quenching parameter'' $\hat{q}$.
The available suite of jet quenching measurements make $\hat q$
one of the experimentally best constrained properties of the hot fluid produced in RHIC
collisions.
We observe that $\hat q$ can be given a model-independent, nonperturbative, quantum field
theoretic definition in terms of the short-distance behavior of a particular light-like
Wilson loop. We then use the AdS/CFT correspondence to obtain a strong-coupling
calculation of $\hat q$ in hot  ${\cal N}=4$
supersymmetric QCD, finding
$\hat{q}_{\rm SYM} =  26.69 \, \sqrt{\alpha_{\rm SYM} N_c} \, T^3$ in the limit
in which both $N_c$ and $4\pi\alpha_{\rm SYM} N_c$ are large.
We thus learn that at strong coupling $\hat q$ is not proportional to the entropy
density $s$, or to some  ``number density of scatterers''
since, unlike the number of degrees of freedom, $\hat q$ does not grow like $N_c^2$.
\end{abstract}
\maketitle
 \vskip 0.3cm


Ultrarelativistic nucleus-nucleus collisions are studied at RHIC and at the LHC to determine
the properties of QCD matter at extreme energy density and temperature~\cite{RHIC,LHC}. 
If we could do the gedanken experiment of deep inelastic scattering on the hot fluid
produced in a heavy ion collision, we could learn a lot.  Even though the short lifetime
of the transient dense state precludes the use of such external probes, a
conceptually similar method is available at RHIC and LHC energies.  This method
is based upon internally generated probes: energetic partons produced
in rare high transverse momentum elementary interactions in the initial stage of the collision, 
which then
interact strongly with the hot, dense fluid produced in the collision as they plough
through it~\cite{jetquenchrev}. The
characterization of  the resulting medium-induced modification of high-$p_T$
parton fragmentation (``jet quenching'') and its connection to
properties of the hot, dense matter that is the object of study
have become one of the most active 
areas of research stimulated by RHIC data~\cite{jetquenchrev}. 
Models which supplement the standard perturbative QCD
formalism for high-$p_T$ hadron production with medium-induced parton energy loss
successfully account for the strong (up to a factor $\sim 5$) suppression of hadronic spectra
in $\sqrt{s_{NN}}=200$ GeV Au-Au collisions at RHIC, its dependence on centrality
and orientation with respect to the reaction plane, and the corresponding reduction of
back-to-back hadron correlations~\cite{Eskola:2004cr,Dainese:2004te}. These models 
typically involve one medium-sensitive
``jet quenching parameter'' denoted $\hat q$.  
This parameter is usually defined only perturbatively,
and is often thought of as proportional to $1/(\lambda^2_D \lambda_{\rm MFP})$,
with $\lambda_D$ the Debye screening length and $\lambda_{\rm MFP}$ some 
perturbatively defined transport mean-free path~\cite{Baier:1996sk}.  
In the present paper, we address the question of how $\hat q$  
can be defined and calculated from first principles in nonperturbative 
quantum field theory, without assuming the existence of quasiparticles with a 
well-defined mean-free path.

There are many indications from data at RHIC and from calculations
of lattice-discretized QCD thermodynamics
that the quark-gluon plasma at temperatures 
not far above the crossover from the hadronic phase 
is strongly coupled. (For example, lattice calculations show that $J/\Psi$ mesons
remain bound~\cite{Asakawa:2003re}; 
for example, there is qualitative agreement between the degree of azimuthal
anisotropy in collisions with nonzero impact parameter seen in RHIC data
and in calculations assuming ideal, zero-viscosity, 
hydrodynamics~\cite{RHIC,Kolb:2003dz,Teaney:2003kp,Shuryak:2004cy}.)
Lattice QCD is the prime example of 
a rigorous calculational method applicable in a hot, strongly coupled, gauge theory.
Because it is formulated in Euclidean space, it is well-suited to calculating static thermodynamic
quantities and less well-suited to calculating transport coefficients,
or dynamical processes of any sort.  It therefore cannot address
parton energy loss itself, and cannot be used to calculate $\hat q$, the property of
the medium that parton energy loss ``measures''. Complementary 
nonperturbative techniques are thus desirable. One such technique is the AdS/CFT
correspondence, which maps nonperturbative problems in certain hot 
strongly coupled gauge theories onto calculable problems
in a dual gravity theory~\cite{AdS/CFT}. This method has been used to calculate the shear 
viscosity in several supersymmetric gauge 
theories~\cite{Policastro:2001yc,Buchel:2003ah,Kovtun:2003wp,Buchel:2003tz,Buchel:2004di,Son:2006em}, as well as for certain diffusion constants~\cite{Policastro:2002se} and
thermal spectral functions~\cite{Teaney:2006nc}. (See Refs.~\cite{Sin:2004yx} for work towards
a dual description of dynamics in heavy ion collisions themselves.) The best-studied example
is the calculation of the shear viscosity $\eta$, where 
the dimensionless ratio $\eta/s$, with $s$ the entropy density,
takes on the value $1/4\pi$ in the large 
number of colors ($N_c$), large 't Hooft coupling ($\lambda\equiv g_{\rm YM}^2 N_c$)
limit of any gauge theory that admits a holographically dual supergravity
description, making this result ``universal''~\cite{Buchel:2003tz}.   
Furthermore, the leading $1/\lambda$ corrections are known~\cite{Buchel:2004di}.
The results thus obtained in supersymmetric 
Yang-Mills theories have been argued 
to be relevant to the QCD matter produced at 
RHIC~\cite{Shuryak:2004cy}. 

In this paper, we shall demonstrate that
the problem of calculating the jet quenching parameter also lends itself to a simple reformulation
in theories with a gravity dual.  After formulating a nonperturbative definition of $\hat q$, 
we shall calculate $\hat q$ in ${\cal N}=4$ supersymmetric
Yang-Mills theory, in the large-$N_c$ and large-$\lambda$ limit.  
We shall close with a comparison to RHIC data, but at present we make
no conjecture for the ``universality'' of our result.  

We begin with the gedanken experiment of DIS on hot matter in thermal equilibrium. 
A virtual photon $\gamma^*$, emitted by the electron, interacts with this matter.
In the rest frame of the thermal matter, the incident virtual photon can be viewed
as a superposition of its hadronic Fock states $\vert \gamma^*\rangle =
\vert q({\bf x})\, \bar{q}({\bf y})\rangle + O(g)$. The lowest lying Fock state
is a  color singlet quark-antiquark dipole whose transverse size in a configuration
space picture is proportional to the inverse of the virtuality of the scattering process 
in momentum space.
In the high-energy limit, the scattering $\hat S$-matrix
can be written in terms of eikonal phase factors $W$, which account for the
precession of the partonic projectile in the color field of the medium~\cite{Kovner:2001vi}. 
For instance, for an incident quark of initial color $\alpha$ and transverse position ${\bf x}$, 
${\hat S}\vert \alpha({\bf x})\rangle = W_{\alpha\, \beta}^F({\bf x})\, \vert \beta({\bf x})\rangle$.
Here 
\begin{equation}
         W({\bf x}) = {\cal P} \exp \left[i \int dx^- A^+({\bf x},x^-) \right]
    \label{eq1}
\end{equation}
is the Wilson line in the representation of the projectile. The quark $\alpha$ propagates 
at fixed transverse position ${\bf x}$ along the light-cone direction $x^-$ through the medium,
which is described by the target color field $A^+$. Then, the ``photoabsorption'' cross section 
for interaction of the $|\bar{q}\, q\rangle$ Fock state of the virtual photon with the target is
\begin{equation}
\sigma^{DIS}=\int d^2{\bf x}\,d^2{\bf y}\,
\psi({\bf x- y})\,\psi^*({\bf x- y})\,
P^{q\bar q}_{\rm tot}({\bf x},{\bf y})\, ,
\label{eq2}
\end{equation}
where $\psi({\bf x- y})\,\psi^*({\bf x- y})$ determines the probability distribution of
color singlet dipoles of transverse size $\vert {\bf x- y}\vert$. The interaction probability
 is~\cite{Kovner:2001vi}
\begin{eqnarray}
  P_{\rm tot}^{q\bar{q}} &=& \left\langle  2-\frac{2}{N_c} 
    {\rm Tr}\left[ W^F({\bf x})\, W^{F\dagger}({\bf y})\right] 
  \right\rangle\ .
  \label{eq3}
\end{eqnarray}
For a color dipole with transverse size $L \equiv \vert {\bf x}-{\bf y}\vert$ which propagates
along the light-cone through a medium of length $L^-$, the only information about
the medium that
enters the calculation of the photoabsorption cross section (\ref{eq2}) is then encoded in the
thermal expectation value of a closed light-like Wilson loop $\langle{W^F({\cal C})}\rangle$, 
whose contour ${\cal C}$ is a rectangle with large extension $L^-$ in the $x^-$-direction 
and small extension $L$ in a transverse direction.
Calculating $\langle{W^F({\cal C})}\rangle$ gives the entire nonperturbative
input to the virtual photoabsorption cross-section (\ref{eq2}).

Inspection of the calculation of the 
energy distribution of the 
medium-induced gluon radiation, the fundamental quantity of interest in
jet quenching calculations, reveals that the medium-dependence enters this calculation
via the expectation value of
a Wilson loop in the adjoint representation with the same contour ${\cal C}$~\cite{Wiedemann:2000za}.
The physical origin of $\langle{W^A({\cal C})}\rangle$ in this
calculation is more difficult to picture than was the case in the DIS example. Here,
the Wilson loop arises
in a configuration space formulation of the gluon radiation cross-section from the combination 
of the gluon emitted at transverse position ${\bf x}$ in the radiation amplitude, with the gluon at 
position ${\bf y}$ in the complex conjugate amplitude. The difference $L = \vert{\bf x}-{\bf y}\vert$ 
is conjugate to the transverse momentum $k_\perp$ of the emitted gluon.
(See Ref.~\cite{Wiedemann:2000za} for details.) 
In jet quenching calculations, one frequently uses the
so-called dipole approximation~\cite{Zakharov:1997uu}, valid for small transverse distances $L$:
 \begin{eqnarray} 
\langle{W^A({\cal C})}\rangle 
	&\approx& \exp\left[ - \frac{1}{4\sqrt{2}}  \hat{q}\, L^-\, L^2 \right]\, .
\label{eq5}
 \end{eqnarray}
Here, $\hat{q}$ is the ``jet quenching parameter'' introduced
previously 
in Ref.~\cite{Baier:1996sk}. (The factor of $\sqrt{2}$ in the denominator arises because
the light-cone distance $L^-$ is larger than the spatial distance travelled, conventionally
used in (\ref{eq5}), by that factor.)
We simply observe that instead of seeing (\ref{eq5}) as an approximation, we
can use it as a nonperturbative definition: $\frac{1}{4\sqrt{2}}\hat q$ is the coefficient 
of the $L^-\,L^2$ term in $\log \langle W^A({\cal C})\rangle$ at small $L$. 

Consider a high energy parton for which the dominant
energy loss mechanism is the radiation of gluons
with large enough $k_\perp\sim 1/L$ that QCD is weakly coupled at the scale $k_\perp$.
This energy loss process cannot be modelled
in a conformal theory that is strongly coupled at all scales.  However,
the strongly coupled physics at scales proportional to $T$ that characterizes the
medium rather than the probe does
enter the energy loss calculation, only via the quantity $\hat q$.
This, and
the fact that time-averaged values of $\hat{q}$ have been determined in comparison with 
data from RHIC in Refs.~\cite{Eskola:2004cr,Dainese:2004te}, 
motivates the calculation of $\hat q$ in strongly coupled gauge theory plasmas.

We have calculated the thermal expectation value of the light-like Wilson loop (\ref{eq5})
for ${\cal N}=4$ super Yang-Mills theory. This is a conformally invariant theory with two
parameters: the rank of gauge group $N_c$ and the 't Hooft coupling 
$\lambda = g_{\rm YM}^2 N_c$.  Its on-shell field content includes eight bosonic and eight
fermionic degrees of freedom, all in the color adjoint representation.
We begin by evaluating the light-like Wilson
loop in the fundamental representation~\footnote{A generic
Wilson loop in ${\cal N}=4$ super Yang-Mills theory depends on six scalar fields, in addition
to the gauge field~\cite{Rey:1998ik}. 
For a light-like contour, the scalar field terms vanish and
the ${\cal N}=4$ Wilson loop coincides with (\ref{eq1}).}.
According to the AdS/CFT correspondence~\cite{AdS/CFT},
in the large-$N_c$ and large-$\lambda$ limits
the thermal expectation value 
$\langle{W^F({\cal C})}\rangle$
can be calculated using the metric for a 5-dimensional curved
space-time describing a black hole in
anti-deSitter (AdS) space~\cite{Rey:1998ik}.  Calling the fifth dimension $r$,
the black hole horizon is at some  $r=r_0$
and the $(3+1)$-dimensional conformally invariant
field theory itself  ``lives'' at $r\rightarrow \infty$.
The prescription for evaluating $\langle{W^F({\cal C})}\rangle$
is that we must find the extremal action surface in the five-dimensional AdS spacetime
whose boundary at $r\rightarrow\infty$ is the contour
${\cal C}$ in Minkowski space $R^{3,1}$.  $\langle{W^F({\cal C})}\rangle$ is
then given by 
 \begin{equation} 
\langle{W^F({\cal C})}\rangle = \exp\left[ -  S ({\cal C})\right] 
	\, ,
\label{eq4}
 \end{equation}
with $S$ the action of the extremal surface~\cite{Rey:1998ik}, subject
to a suitable subtraction that we discuss below.

Using light-cone coordinates  $x^\mu = (r,x^\pm, x^2,x^3)$, the AdS black hole metric is given 
by~\cite{AdS/CFT}
 \begin{eqnarray}
ds^2 
 & = & - \left({r^2 \over R^2} + f \right) dx^+ dx^- 
 + {r^2 \over R^2} (dx_2^2 + dx_3^2)
 \nonumber \\
 &&+  \frac{1}{2} \left({r^2 \over R^2} - f \right)
 \left((dx^+)^2 + (dx^-)^2 \right) +
{1 \over f} dr^2 \nonumber \\
 & = & G_{\mu \nu} dx^\mu dx^\nu \, ,
 \label{eq6}
 \end{eqnarray}
where $f = {r^2 \over R^2} \left(1 - {r_0^4 \over r^4} \right)$ with $R$ the curvature radius
of the AdS space.
Here, 
the temperature $T$ of the Yang-Mills theory is given by the Hawking temperature of the 
black hole,  $T = \frac{r_0}{ \pi R^2}$, 
and $R$ and the string tension
$1/2\pi\alpha'$ are related to the t'Hooft coupling by $\frac{R^2}{\alpha'} = \sqrt{ \lambda}$.

We parameterize the surface whose action $S({\cal C})$ is to be 
extremized by  $x^\mu = x^\mu (\tau,\sigma)$, 
where $\sigma^\alpha = (\tau, \sigma)$ denote the coordinates parameterizing the
worldsheet. The Nambu-Goto action for the string worldsheet is given by
 \begin{equation} 
S ={1 \over 2 \pi \alpha'} \int d\sigma d \tau \, \sqrt{ \det g_{\alpha
\beta}}
\label{eq7}
 \end{equation}
with $g_{\alpha \beta} = G_{\mu \nu} \partial_\alpha x^\mu \partial_\beta x^\nu$
the induced metric on the worldsheet. This action is invariant under coordinate 
changes of $\sigma^\alpha$, and we can set $\tau = x^-$ and $\sigma = x_2$,
taking the ``short sides'' of ${\cal C}$ with length $L$ along the $x_2$ direction.
Since $L^- \gg L$, we can assume that the surface is
translationally invariant along the $\tau$ direction, i.e. 
$x^\mu=x^\mu (\sigma, \tau)=x^\mu (\sigma)$.
The Wilson loop lies at constant $x_3$ and at constant $x^+$, 
so $x_3 (\sigma) = {\rm const}$, $x^+ (\sigma) = {\rm const}$.
For the bulk coordinate $r$, we implement the requirement
that the world sheet has ${\cal C}$ as its boundary by
imposing $r \left(\pm  {L \over 2} \right) = \infty$, which preserves the symmetry
$r(\sigma) = r(-\sigma)$.  The action (\ref{eq7}) now takes the form
\begin{equation}
 S = {\sqrt{2} \,r_0^2 \, L^- \over 2 \pi \alpha' R^2} \int_0^{{L \over 2}} d \sigma \,
\sqrt{1+{r'^2 R^2 \over f r^2} }\ ,
\label{eq8}
 \end{equation}
where $r'=\partial_\sigma r$. The equation of motion for $r (\sigma)$ is then
 \begin{equation} 
r'^2 = \gamma^2 {r^2 f \over R^2}
\label{eq9}
 \end{equation}
with $\gamma$ an integration constant. Eq.~(\ref{eq9}) has two
solutions. One has $\gamma = 0$ and hence $r'=0$, meaning $r(\sigma)=\infty$
for all $\sigma$: the surface stays at infinity. This solution is not of interest. 
The other solution has $\gamma>0$. It ``descends'' from $r(\pm \frac{L}{2})=\infty$ 
and has a turning point where $r'=0$ which, by symmetry, must occur at $\sigma=0$.
From (\ref{eq9}), the turning point must occur where $f=0$, implying $r'=0$ at 
the horizon $r=r_0$.  The surface descends from infinity, skims the horizon, and
returns to infinity.  Note that the surface descends all the
way to the horizon regardless of how small $L$ is~\footnote{We have checked that the surface 
also 
has this qualitative shape, with $r=r_0$ at $\sigma=0$ for any $L$, in the ${\cal N}=2^*$  theory
analyzed in 
Ref.~\cite{Buchel:2003ah} which is a deformation of the ${\cal N}=4$ theory in which some 
fields are given masses that we take as
small compared to $T$.  This demonstrates
that this feature of the solution is not specific
to ${\cal N}=4$.}.
This is reasonable
on physical grounds, as we expect $\hat q$ to describe the thermal medium rather than
ultraviolet physics.  Knowing that $r=r_0$ at $\sigma=0$,
(\ref{eq9}) can be integrated, yielding
\begin{equation}
\frac{L}{2}=\frac{R^2}{\gamma}\int_{r_0}^\infty \frac{dr}{\sqrt{r^4-r_0^4} }
= \frac{a R^2}{\gamma r_0}
\end{equation}
where $a=\sqrt{\pi}\Gamma(\frac{5}{4})/\Gamma(\frac{3}{4})\approx 1.311$.
We then find
\vspace{-0.02in}
\begin{equation}
S   = \frac{\pi \sqrt{\lambda} L^- L T^2}{2\sqrt{2}} \sqrt{1 + {4 a^2 \over
 \pi^2 T^2  L^2}}\  ,
  \label{eq10}
 \end{equation}
\vspace{-0.02in}
where we have used $r_0=\pi R^2 T$ and $\frac{R^2}{\alpha'} = \sqrt{\lambda}$.

Recalling our discussion of the DIS example, we now note that whereas 
we wish to evaluate
the interaction of a ``bare'' $|q \bar q\rangle$ Fock state of the virtual photon
with the medium, what we have calculated above includes the ``self-energy''
of the high energy quark and antiquark, moving through the medium as if in the 
absence of each other.
We perform the required subtraction upon noting that in addition to the
extremal surface constructed above, there is another trivial one given by
two disconnected worldsheets, each of which descend from $r=\infty$
to $r=r_0$ at constant $x_2$, one
at $x_2=+\frac{L}{2}$ the other at $x_2=-\frac{L}{2}$.
The total action for these two surfaces is
 \begin{equation}
 S_0 = {2 L^- \over 2 \pi \alpha'}  \int_{r_0}^\infty dr \, \sqrt{g_{--}
 g_{rr}} = \frac{a \sqrt{\lambda} L^- T}{\sqrt{2}} \, .
 \label{eq11}
 \end{equation}
Subtracting this self-energy, we obtain
\begin{equation}
S_I \equiv S-S_0\ ,
\end{equation}
the $L$-dependent interaction between the quark pair moving through the medium.
Because the quarks
are moving at the speed of light, both $S_0$ and $S$ are finite:  no ultraviolet
subtraction like that required for static quarks is needed. 
$\exp[-S_I]$ is the thermal expectation value of the Wilson loop (\ref{eq4})
in the fundamental representation, and would be our final result were
we actually interested in DIS. 
Note that $S_I$ vanishes for $L\rightarrow 0$ as it must, since the photoabsorption
probability (\ref{eq3}) vanishes in that limit.

We now evaluate the adjoint Wilson loop (\ref{eq5}) that determines $\hat q$.
For $SU(N_c)$, the Wilson line
in the adjoint representation 
can be obtained using the identity  
${\rm Tr} W = {\rm tr} W {\rm tr} W^\dagger - 1$, where ${\rm Tr}$ and ${\rm tr}$ 
denote traces in the adjoint and fundamental representations, respectively.
This implies that in the large-$N_c$ limit, the adjoint Wilson loop (\ref{eq5}) is
given by $\exp\left[ - 2S_I\right]$. 
The jet quenching parameter
$\hat q$ in (\ref{eq5}) is determined by 
the behavior of $2 S_I$ for small transverse
distances, $L T \ll 1$. We find
\vspace{-0.02in}
 \begin{equation} 
 \label{eq13}
2 S_I =   \frac{\pi^2}{4\sqrt{2}\, a}\sqrt{\lambda} \,T^3  L^- L^2  + {\cal O}(T^5 L^- L^4)\, 
 \end{equation}
 \vspace{-0.02in}
and can now use (\ref{eq5}) to read off our central result:
\begin{equation}
{\hat q}_{\rm SYM} = \frac{\pi^2}{a}\sqrt{\lambda}\,T^3 = \frac{\pi^{3/2}\Gamma(\frac{3}{4})}
{\Gamma(\frac{5}{4})}\sqrt{\lambda}\,T^3 \approx 26.69 \sqrt{\alpha_{\rm SYM}N_c} \,T^3
\label{eq14}
\end{equation}
in  ${\cal N}=4$ supersymmetric Yang-Mills theory in the large-$N_c$ and large-$\lambda$ 
limits.

Perhaps the most striking qualitative feature of our result
is that at strong coupling 
$\hat q$ is proportional to $\sqrt{\lambda}$,
not to the number of degrees of freedom $\sim N_c^2$. 
This means that at strong coupling $\hat q$ cannot be thought of as ``measuring'' 
either the entropy density $s$ or what is sometimes described as a ``gluon number density''
or $\varepsilon^{3/4}$ as had been expected~\cite{jetquenchrev},
since both $s$ and the energy density $\varepsilon$
are proportional to $N_c^2 \lambda^0$.  
Instead, it appears that $\hat q$ is
better thought of as a measure of $T^3$\footnote{This conclusion can be strengthened upon
redoing our analysis for the $(p+1)$-dimensional super-Yang-Mills theories (with 16
supercharges) living at the boundary of the geometry
describing a large number of non-extremal black Dp-branes~\cite{Itzhaki:1998dd}.  
In these theories, in
which $\lambda$ has mass dimension $(3-p)$, we find 
$\hat q \propto T^2 (T\sqrt{\lambda})^{2/(5-p)}$ whereas 
$s\propto N_c^2 \lambda^{(p-3)/(5-p)} T^{(9-p)/(5-p)}$ and $\varepsilon\propto Ts$.
This means that for $p\neq 3$, $\hat q$, $s$ and $\varepsilon^{3/4}$
have different $T$-dependence, not only different $N_c$-dependence.}.

We would be remiss not to attempt a comparison to implications
of RHIC data.  Taking $N_c=3$ and  $\alpha_{\rm SYM}=\frac{1}{2}$, reasonable
for temperatures not far above the QCD phase transition,  we shall use $\lambda=6\pi$
to make estimates. From (\ref{eq14}), we find $\hat q=$4.5, 10.6, 20.7 GeV$^2/$fm
for $T=$ 300, 400, 500 MeV.  In a heavy ion collision, $\hat q$ decreases with time $\tau$
as the hot fluid expands and cools. The time-averaged $\hat q$ which has been determined in
comparison with RHIC data is 
$\overline{\hat{q}} \equiv \frac{4}{(L^{-})^2} \int_{\tau_0}^{\tau_0+ L^-/\sqrt{2}} \tau\, \hat{q}(\tau)\ d\tau$,
found to be around 5-15~GeV$^2/$fm~\cite{Eskola:2004cr,Dainese:2004te}. If we
assume a one-dimensional Bjorken
expansion with $T(\tau) = T_0 \left(\frac{\tau_0}{\tau}\right)^{1/3}$, 
take $\tau_0=0.5$~fm,
and take $L^-/\sqrt{2}=2$~fm, the estimated mean distance travelled in the
medium by those hard partons which ``escape'' and are detected~\cite{Dainese:2004te}, 
we find that to obtain $\overline{\hat{q}}=5$~GeV$^2/$fm from (\ref{eq14})
we need $T_0$ such that $T(1~{\rm fm})\approx 310$~MeV, only slightly higher than
expected~\cite{Kolb:2003dz}. 
Equivalently, the $\hat q$ we find from (\ref{eq14}) is 
slightly smaller than that suggested by RHIC data.
First, this could indicate that in going from hot ${\cal N}=4$ to the QCD quark-gluon 
plasma, as the number of color adjoint degrees of freedom is decreased by a factor of $2/15$
(and the color fundamental quarks are added) the jet quenching parameter 
{\it increases} slightly.  Second, it could indicate that the $\overline{\hat q}$ extracted
from data is somewhat high, either 
because
the medium through which the 
energetic parton moves has a flow velocity transverse
to the parton's motion
or because additional sources of energy loss 
are also significant, for example collisions without gluon radiation or processes 
occurring before
the medium is in local thermal equilibrium. 
Testing the second conclusion likely requires further experimental data;
the first can be tested by evaluating $\hat q$ in other gauge theories
with gravity duals.  Unless our result turns out to be ``universal'', 
which would be interesting in its own right, such calculations
would indicate the direction in which $\hat q$ changes as the theory
is made more QCD-like.
Calculating the $1/\lambda$ corrections to our result 
would also be informative.

%

\begin{acknowledgments}

We thank J. Maldacena, B. M\"uller and C. Nunez for discussions.
HL is supported
in part by the A.~P.~Sloan Foundation and the U.S. Department
of Energy (DOE) OJI program.  This research was
supported in part by 
the DOE under contracts
\#DE-AC02-05CH11231, \#DE-AC02-98CH10886
and 
\#DF-FC02-94ER40818.

\end{acknowledgments}

%

\end{document}